\documentclass{XrU2005}
\usepackage{graphicx}
\usepackage{epsfig, subfigure} 
\title{Searching for Sharp Surface Brightness Discontinuities: 
a Systematic Study of Cold Fronts in Galaxy Clusters}
\author{S. Ghizzardi, S. Molendi, A. Leccardi, M. Rossetti}
\affil{IASF Milano, Istituto Nazionale di Astrofisica, Milano, Italy}

\begin{document}


\maketitle

\begin{abstract}
We perform a systematic search of cold fronts in a sample of 62 clusters observed with 
XMM-Newton with redshift ranging from 0.01 to 0.3. 
We detect one or more cold fronts in 21 (34\%) of our objects.
A large fraction (87.5\%) of nearby clusters $0.01 < z < 0.04$ host a 
cold front while only 20\% of the distant clusters, mostly 
merging clusters, do so. The absence of sharp surface brightness 
discontinuites in distant cool cores is most likely a consequence of the insufficent 
spatial resolution of our images.
Some nearby cool core clusters show 
a dislocation between the surface brightness and the pressure 
peak. This implies that the cool central gas is displaced from the 
bottom of the gravitational potential well and likely sloshing.
\end{abstract}

\section{Introduction}

The high spatial resolution of the Chandra X-ray telescope has led to the discovery of 
several phenomena within galaxy clusters.
In particular, Chandra observations revealed the existence
of very sharp discontinuities  in the X-ray surface brightness of several 
clusters.
The drop in the X-ray surface brightness  
is accompanied by a jump of similar magnitude in the gas temperature,
so that the pressure does not change drastically across the front. 
This feature has been named ``cold front'' \citep{Vik:2001}.

Cold fronts have been initially observed in merging clusters. 
The prototype cold fronts are hosted in A2142 \citep{Maxim:2000},  
A3667 \citep{Vik:2001} and 1E0657-56 \citep{Maxim:2002}. In these clusters, 
the cold front delineates the edge of the cool core of the merging 
substructures
which have survived the merger and is rapidly moving throughout the 
shock-heated ambient gas \citep{Maxim:2000}.

Cold fronts have been detected also in the core of some relaxed clusters (e.g. 
A1795: \citealp{Maxim:2001}; RX J1720.1+2638:
\citealp{Mazzotta:2001}; A496: \citealp{Dupke:2003}; 2A 0335+096: \citealp{Mazzotta:2003}).
The presence of cold fronts in cool cores provides evidence of gas 
motions and possiblly of departures from hydrostatic equilibrium.
Cold fronts represent a unique tool of investigation of the internal 
dynamics of clusters, especially in the regrettable absence of instrumentation 
capable of a direct detection of gas motions.
To study and characterize cold fronts, we built a sample of 62 clusters observed 
with XMM-Newton and carried out a systematic search for 
surface brightness and temperature discontinuities. We studied the occurrence of cold 
fronts in different redshift ranges and for different types of clusters (merging and cool 
cores). Particular attention has been devoted to cold fronts in relaxed clusters.  

The sample is described in \S \ref{sec:sample}. Technical details concerning the procedure 
used for searching cold fronts are given in  \S \ref{sec:search}.
In \S \ref{sec:freq}, we discuss the occurrence of cold fronts in clusters. 
The nature of cold fronts in cool cores has been discussed in \S \ref{sec:CC}. Finally, in 
\S \ref{sec:summary} we summarize our results.

\section{The data sample}
\label{sec:sample}
We have performed a systematic search and characterization of cold fronts from 
a large sample of clusters observed with XMM-Newton. The large collecting area of the 
EPIC telescope onboard the XMM-Newton satellite 
allows a detailed inspection of the spectral properties of the galaxy clusters, which 
are important to study the dynamics of the core.
Namely, we have selected a sample of 62 clusters of galaxies. The sample includes two 
different subsamples.
The first comprises roughly 20 nearby bright clusters 
with redshifts in the range $[0.01-0.1]$. The second subsample 
comprises all the clusters available in the XMM-Newton public archive
up to March 2005,  having redshifts in the range $[0.1-0.3]$ \citep{Leccardi:2005}.

\begin{figure}
\centering
\epsfig{file=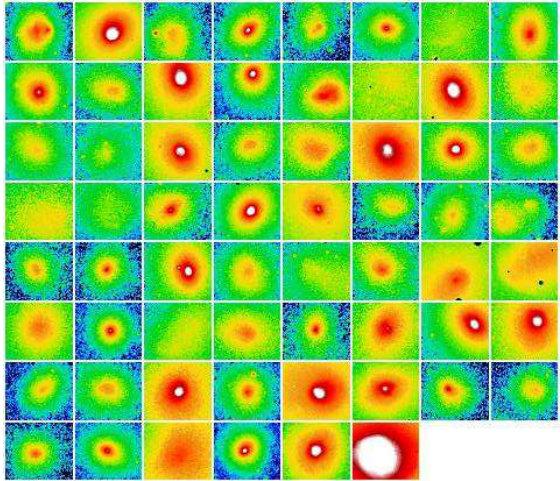,width=0.8\linewidth,angle=270}
\caption{The EPIC surface brightness images for the 62 clusters in our sample. 
Flux is in units $ 10^{-15} {\rm erg} {~\rm cm}^{-2} 
{\rm s}^{-1} {\rm pixel}^{-1}$ .}
\label{fig:allmaps}
\end{figure}

The sample (see Fig.~\ref{fig:allmaps}) includes a wide variety of clusters, 
both merging and cool cores.
It is worth noting that we did not fix an objective selection parameter to build our 
sample.
However, the criteria used for the selection of the 
clusters are not related to the existence of the cold fronts. 
Hence, no bias is {\sl a priori} present in our sample as far as cold fronts occurrence is 
concerned and for our purposes this sample is sufficiently representative of the cluster 
population up to redshift 0.3.

\section{Searching for cold fronts.}
\label{sec:search}
To find cold fronts and to study their properties, we need accurate surface brightness
and temperature maps and profiles. 
Profiles should be determined in different directions for each cluster.
For clusters with good statistics, we have used the algorithm developed 
by the Milano group to build the maps.
The procedure consists in binning the cluster images,
using the adaptive binning algorithm
developed by \citet{CC:2003} and based on the Voronoi tessellation technique.
The temperature ($T$) and the surface brightness ($\Sigma$) values in each bin are 
determined 
through a broad band fitting method \citep{Rossetti:2005}. Errors are determined 
for both $\Sigma$ and $T$ in each bin.

\begin{figure}
\centering
\epsfig{file=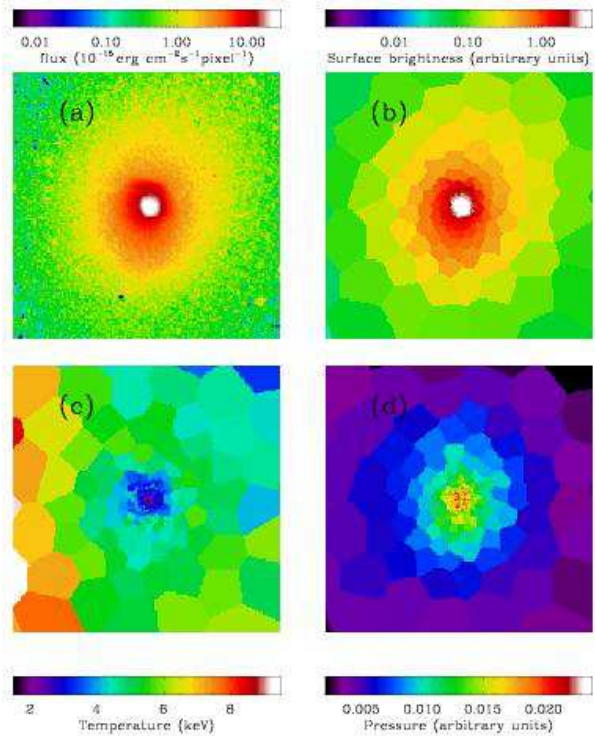,width=0.99\linewidth,angle=0}
\caption{(a) EPIC surface brightness image for A496. (b) Surface brightness map, (c) 
temperature map and (d) pseudo-pressure map for A496.}
\label{fig:adbin}
\end{figure}

As an example, in Fig.~\ref{fig:adbin} we show the maps obtained 
applying our procedure to the XMM-Newton observation of the 
galaxy cluster A496.
In Fig. \ref{fig:adbin}(a) and (b), we report respectively the EPIC (MOS + PN) surface 
brightness image and the surface brightness map derived using our procedure. 
In Fig. \ref{fig:adbin}(c), the A496 temperature map is shown; the central cool core is 
clearly visible. 
The quality of the A496 
observation is quite good, so the derived maps are very accurate and 
the surface brightness map is very similar to the surface brightness image.

\begin{figure*}
\begin{center}
\epsfig{file=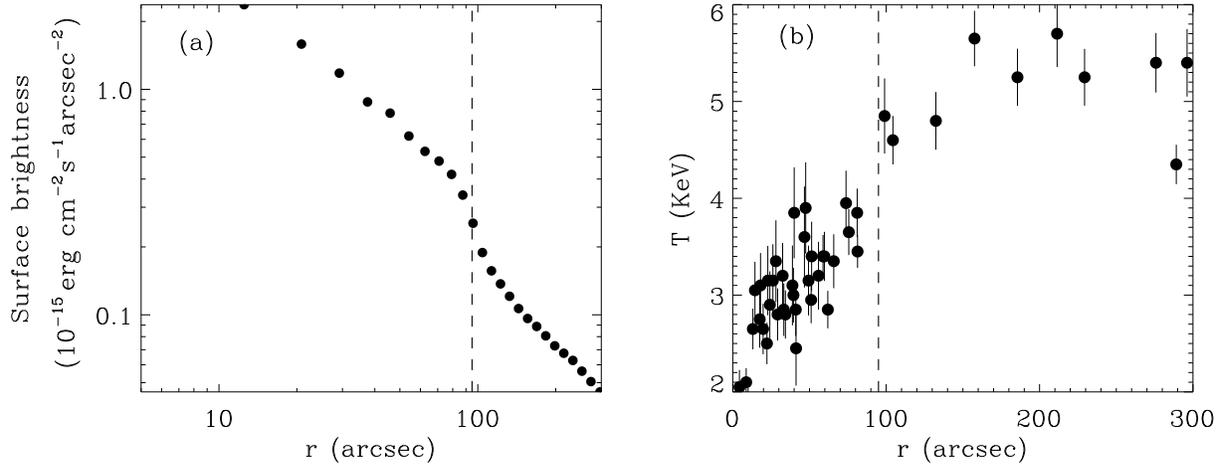,width=0.4\linewidth,angle=90}
\caption{(a) Surface brightness profile and (b) temperature profile for A496. The profiles 
have been derived in the sector $60^{\circ}-120^{\circ}$. The dashed line marks the cold 
front position, $\sim 90$ arcsec from the peak. }
\label{fig:A496_prof}
\end{center}
\end{figure*}

Once the surface brightness and the temperature have been obtained, 
other thermodynamic quantities (pressure and entropy) can be derived.  Although a 
deprojection procedure should be applied to derive these quantities, we work in the 
approximation that the surface brightness is $ \Sigma \sim n^2 $, where $n$ 
is the electronic density. The (pseudo)-pressure and the (pseudo)-entropy are 
correspondingly derived through the projected quantities
$P = \Sigma ^{1/2} T$ and $S = T/{ \Sigma ^{1/3}}$.
The (pseudo)-pressure map for A496 is reported in  Fig.~\ref{fig:adbin}(d). 
A cold front at roughly 90 arcsec from the peak in the N-NW sector 
(roughly $60^{\circ} - 120^{\circ}$)
is clearly observed in the surface brightness and temperature maps.
The $\Sigma$ and $T$ (see Fig.~\ref{fig:A496_prof}) profiles in this sector show that the 
surface brightness has a sharp discontinuity (approximatively by a factor of 2-3) 
and that the temperature across the edge varies by a factor of 1.5.
Correspondingly the pressure does not show an abrupt drop and there is 
approximate pressure equilibrium across the front.

\begin{figure}
\centering
\epsfig{file=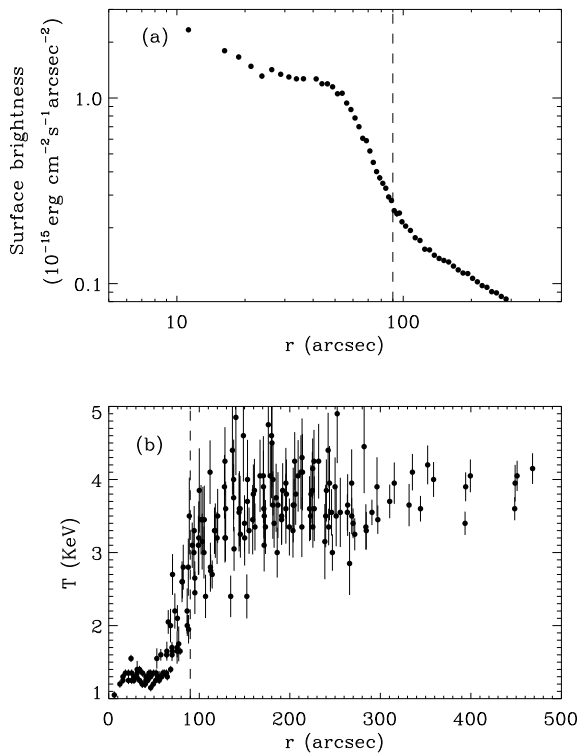,width=0.99\linewidth,angle=0}
\caption{a) Surface brightness profile and (b) temperature profile for Centaurus. The 
profiles 
have been derived in the sector $120^{\circ}-150^{\circ}$. The dashed line marks the cold 
front position, $\sim 90$ arcsec from the peak.}
\label{fig:Cent_prof}
\end{figure}

Another cluster with high statistics is Centaurus.  
The surface brightness and the temperature profiles 
(in Fig.~\ref{fig:Cent_prof} (a) and (b) respectively) of the NE sector ($\sim 
120^{\circ} - 150^{\circ}$) show that Centaurus hosts a cold front at $\sim 90$ arcsec
from the peak. The surface brightness drops by a factor of 4-5 and the temperature 
rises by almost a factor of 3. Also in this case the pressure is almost constant 
across the front.

\begin{figure}
\begin{center}
\epsfig{file=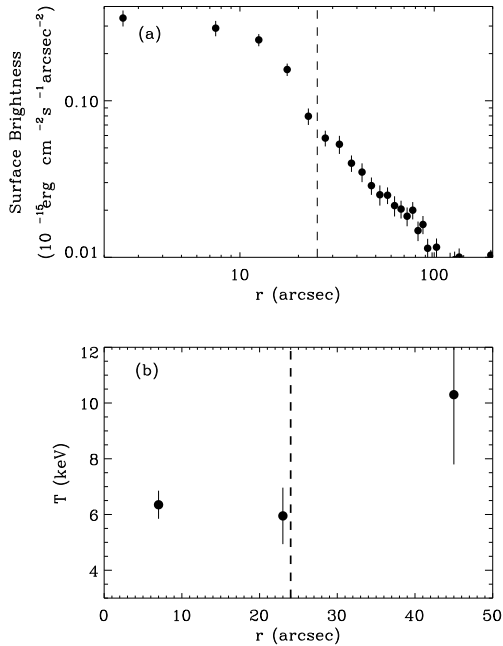,width=0.85\linewidth,angle=0}
\caption{ (a) Surface brightness profile and (b) temperature profile for A1300. The 
temperature profile 
have been derived applying the broad band fitting procedure to manually selected macro 
regions. The dashed line marks the cold front 
position, $\sim 25$ arcsec from the peak.}
\label{fig:A1300}
\end{center}
\end{figure}

For clusters having poor statistics the adaptive binning procedure does not allow us to 
produce detailed maps. Hence, for these clusters the broad band fitting method has been 
applied to manually selected macro regions. Regions are built in such a way as 
to follow the cold 
front feature so that the temperature profile across the cold front can be determined.
This is the case of A1300, where a discontinuity in the surface brightness (see 
Fig.~\ref{fig:A1300}a) is detected at $\sim 25$ arcsec from the peak (in the W-NW 
direction).
The temperature profile (Fig.~\ref{fig:A1300}b) is not as detailed as in the 
previous cases, nevertheless there is an indication of a temperature rise.

\section{Occurrence of Cold Fronts in galaxy clusters}
\label{sec:freq}
The systematic analysis of the surface brightness of the clusters of our sample 
brought to the detection of cold fronts in 21 objects (and probably also in 
other 4 clusters where the presence of a cold front is not clear and needs some further 
investigation).
This corresponds to a percentage of 34\% (40\% if the uncertain cases are also 
accounted) in the redshift range [0.01-0.3].
It is interesting to study also the frequency of cold fronts in different redshift ranges.
Our analysis shows that if we progressively reduce the sample, excluding gradually the 
more distant clusters, the fraction of 
clusters having a cold front increases. In particular, in the subsample of the nearby 
clusters (with redshift in the [0.01-0.04] range), 87.5\% of our objects
exhibit one or more cold fronts.
Considering that projection effects can hide a non-negligible fraction of 
cold fronts, such frequency implies that probably all the nearby 
clusters host one or more cold fronts.
This result is in agreement with \citet{Maxim:aph} who analyzed a sample of 37 relaxed 
nearby clusters observed with {\sl Chandra} and find that roughly 70\% of the nearby cool 
core clusters of their sample host a cold front \citep{Maxim:aph}.

While cold fronts are a common feature in nearby clusters, either merging or cool 
cores, for distant clusters their frequency is different for different types of clusters. 
For clusters with redshift larger than 0.1, cold fronts are mostly detected in merging 
clusters, while only few cool core clusters exhibit one.
Since cold fronts in cool cores are in general quite near to the peak and less 
prominent than in merging clusters,
it is likely that the resolution of XMM-Newton is not enough to 
detect sharp discontinuities for this class of clusters at high redshifts. 

\section{The cold fronts in relaxed galaxy clusters}
\label{sec:CC}
Our sample includes a large number of nearby relaxed clusters and we have 
already shown that most of them host a cold front. 
The presence of a cold front in the center of cool core clusters is a clear 
indication that the ICM in these objects is not completely static.
Several Chandra observations of clusters have revealed that 
relaxed clusters often have disturbed cores with complex morphological structure.
Two major pictures have been proposed for cold 
fronts in cool core clusters. \citet{Maxim:aph} proposed that the central 
cool gas is sloshing in the underlying gravitational potential well. In this 
scenario, the cold front is the boundary between the displaced cooler gas and the ambient 
hot gas.
The gas could have been displaced by some AGN central activity or as a consequence of 
some past minor merging process \citep{Maxim:aph,Maxim:2001}.

An alternative picture has been recently proposed by \citet{TH:2005}.
Using numerical simulations, the authors show how the oscillations of the dark matter (and 
correspondingly of the gravitational potential well) can produce cold 
fronts. 
In this scenario, it is the potential well that is oscillating and not the gas within it;
the gas and the dark matter are oscillating together, since dark matter 
movement induces gas 
motions. The resulting compression of the isodensity contours along the oscillations 
generates the cold front feature.

A detailed analysis of cold fronts in relaxed clusters can help us improve 
our understanding of the nature of the phenomenon. 
We selected from our sample some cool core clusters.
For most of them, the derived temperature and (pseudo)-pressure maps
are detailed and allow a good inspection of the cool core physics.

\begin{figure*}
\centering
\epsfig{file=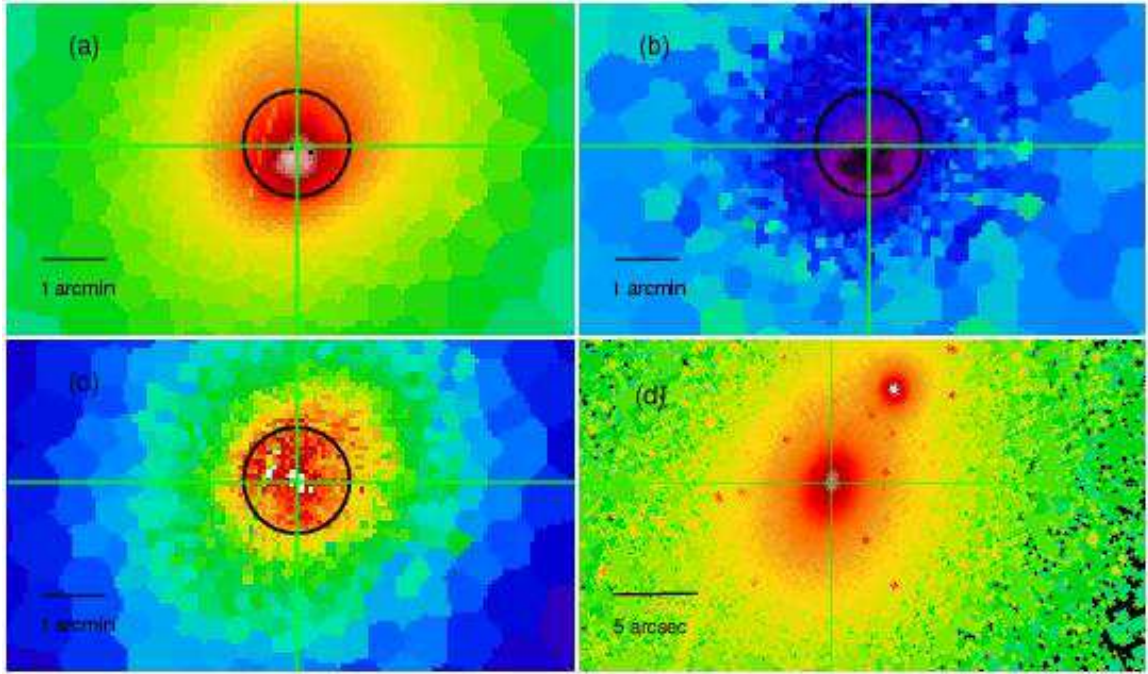,width=0.9\linewidth,angle=0}
\caption{(a) Surface brightness (b) temperature and (c) pseudo-pressure maps for 
the core of  2A0335+096. The cold front is observed in the S-SW sector $\sim 
50$ arcsec fron the peak. (d) Optical HST image of the central cD galaxy of 
2A0335+096. Crosshairs are centered on the cD position. The circle approximates an 
isocontour in the pressure map. The pseudo-pressure peak matches the cD galaxy position, 
while the surface brightness and the temperature peaks are displaced in the direction of 
the cold front.}
\label{fig:2A0335_peak}
\end{figure*}

\begin{figure}
\centering
\epsfig{file=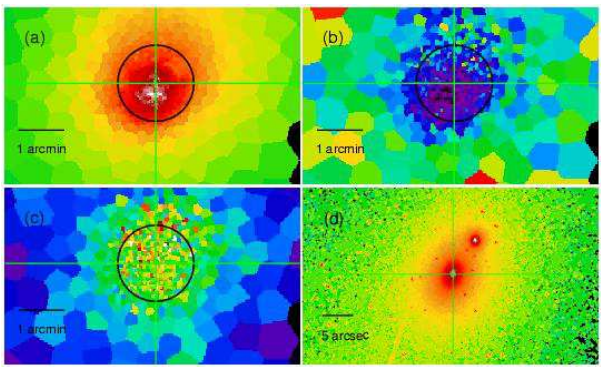,width=1.\linewidth,angle=0}
\caption{Maps for 2A0335+096 obtained from the analysis of a Chandra observation. Panels 
are as in Fig.~\ref{fig:2A0335_peak}. The displacement of the surface brightness peak in 
the direction of the cold front is even more evident here.}
\label{fig:Chandra_2A}
\end{figure}

2A 0335+096 is a cool core cluster with a cold front in the S-SW sector $\sim 
60$ arcsec from the peak \citep[see also][]{Mazzotta:2003}.
In Fig.~\ref{fig:2A0335_peak} we show the maps obtained applying our algorithm to the 
XMM-Newton data for 2A 0335+096.
In Fig.~\ref{fig:2A0335_peak}(a) and (b), we show respectively the surface brightness and 
the temperature maps for the central region of the cluster.
In Fig.~\ref{fig:2A0335_peak}(c) the (pseudo)-pressure is reported. Finally, in 
Fig.~\ref{fig:2A0335_peak}(d) the optical image of the 
central cD galaxy observed with HST is shown. This last image has been repeatedly zoomed 
in, for a better view of the galaxy, while the scales of the three images derived from 
XMM-Newton observation match each other. 
The crosshair is centered on the cD galaxy. The circle is an approximation of an isophote 
of the pressure image. The peak of the 
pressure perfectly matches the cD galaxy position, while the surface brightness peak 
and the temperature minimum are significatively displaced in the direction of the cold 
front (S-SW direction).
The same analysis has been applied also to a {\sl Chandra} observation of 2A 0335+096.
The corresponding maps are reported in Fig.~\ref{fig:Chandra_2A}. The panels are as in 
Fig.~\ref{fig:2A0335_peak}. The displacement between the surface brightness and the 
pressure (or the cD galaxy) peak is even more evident here.

The mismatch between pressure and surface brightness can help improve our understanding 
of cold fronts.
Both the gas pressure and the cD trace the gravitational potential of the cluster, while 
the surface brightness describes the X--ray emitting gas. 
The dislocation of the surface brightness peak in the direction of the cold front,
indicates that the thermal gas is not at 
the bottom of the gravitational potential well, in agreement with a the sloshing scenario 
proposed by \citet{Maxim:aph}.

\begin{figure}
\centering
\epsfig{file=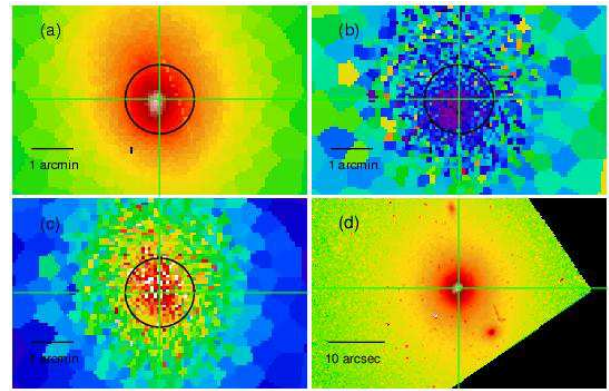,width=1.\linewidth,angle=0}
\caption{Maps for A1795. Panels 
are as in Fig.~\ref{fig:2A0335_peak}. The cold front is in the southern sector $\sim 70$ 
arcsec from the peak. The displacement of the surface brightness peak in 
the direction of the cold front can be observed.}
\label{fig:A1795_peak}
\end{figure}

A1795 (see Fig.~\ref{fig:A1795_peak}) shows a similar mismatch.  
This cluster has a cold front in the southern sector $\sim 70$ arcsec from the peak. 
\citet{Maxim:2001} propose that 
the sloshing gas is now at the maximum displacement from the peak with a zero 
velocity. The 
displacement from the bottom of the potential well is visible in Fig.~\ref{fig:A1795_peak} 
although it is less evident than in 2A 0335+096 maps. As for 2A 0335+096, the 
position of the cD (in the HST  image) and the pressure peak match each other, while the
surface brightness peak is displaced towards the southern direction.
 
\begin{figure}
\centering
\epsfig{file=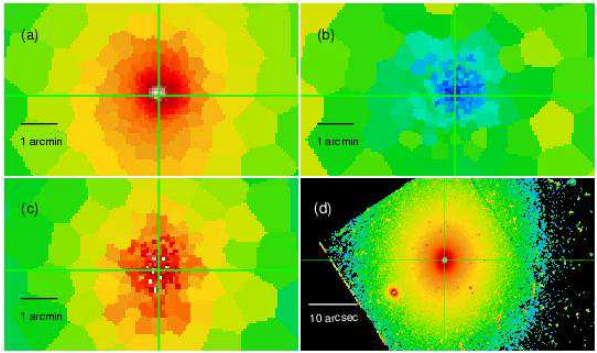,width=0.99\linewidth,angle=0}
\caption{Maps obtained for A496. Panels 
are as in Fig.~\ref{fig:2A0335_peak}. The cold front is in the N-NW sector $\sim 90$ 
arcsec from the peak. The displacement of the surface brightness peak in 
the direction is not observed, at the available resolution of $\sim 2$ kpc.}
\label{fig:A496peak}
\end{figure}

A different picture emerges from the analysis of the cluster A496 (see maps in
 Fig.~\ref{fig:A496peak}).
As already outlined in \S~\ref{sec:search}, A496 is a cool core cluster and exhibits a 
cold front in the N-NW direction.   
Fig.~\ref{fig:A496peak} shows (at the available resolution $\sim 2$ kpc) no displacement 
between the surface brightness and the pressure (or the cD position).
This is compatible with a Tittley-Henriksen 
picture where the dark matter and the ICM oscillate together.
However, the Markevitch et al. picture is not ruled out: the oscillating gas may be 
observed just while passing in the center of the potential well.

At the present time, the analysis of these three clusters leaves room for both the 
scenarios described above.
A systematic analysis of the temperature, pressure and metal maps for all the 
cool core clusters of the sample is needed to
address the nature of this class of cold fronts.

\section{Summary}
\label{sec:summary}

In the absence of instrumentation capable of detecting gas motions, cold 
fronts are our primary tool to study the internal dynamics of the ICM. 
We have selected a large sample including 
62 galaxy clusters observed with XMM-Newton, having a redshift in the [0.01-0.3] range.
The systematic search of surface brightness discontinuities has brought to the detection 
of cold fronts in 21 (34\%) clusters of galaxies. 
Almost all the nearby clusters (either merging and cool cores) host a 
cold front while, as far as distant clusters are concerned, 
cold fronts are mostly detected in 
merging clusters and only few relaxed clusters host one. Probably, cold fronts are not 
detected in distant relaxed clusters because of the insufficient resolution of our 
observations. 

Through the analysis of the surface brightness, temperature, 
(pseudo)-pressure maps for some cool core clusters we have outlined a method which can 
help us improve our understanding of 
this phenomenon. In some clusters, 
(e.g. 2A0335, A1795), the 
surface brightness peak (which describes the X--ray emitting gas) is displaced 
from the pressure (and the central cD) peak which traces the gravitational potential 
well. This is an indication that the thermal gas is displaced from the bottom of the 
potential well and probably is sloshing \citep[see][]{Maxim:aph}.
On the contrary, in A496 all the peaks (surface brightness, pressure and cD galaxy)
match each other. This behavior favors the scenario 
proposed by \citet{TH:2005} where the dark matter itself (carrying the 
gravitational potential and the gas) is oscillating. This is also compatible with  
the \citet{Maxim:aph} picture as the oscillating gas may be 
observed just while passing in the center of the potential well.

A systematic study of the dislocation of the 
surface brightness peak for all the relaxed clusters of our sample 
will help us address this issue.


\begin{thebibliography}{}


\bibitem[Cappellari \& Copin(2003)]{CC:2003} Cappellari M. \& Copin Y. 2003,  \mnras, 342, 
345

\bibitem[Dupke \& White III(2003)]{Dupke:2003} Dupke R. \& White III R.E., 2003, \apj, 
583, L13

\bibitem[Leccardi et al.(2005; in preparation)]{Leccardi:2005} Leccardi A., et al. 2005,  
in preparation

\bibitem[Markevitch et~al.(2000)]{Maxim:2000} Marchevitch M., et~al. 2000, \apj, 541, 542

\bibitem[Markevitch, Vikhlinin \& Mazzotta(2001)]{Maxim:2001} Markevitch M., Vikhlinin A.
\& Mazzotta P. 2001, \apj,  562, L153

\bibitem[Markevitch et~al.(2002)]{Maxim:2002} Markevitch M., Gonzalez A.H., David L., 
Vikhlinin A., Murray S., Forman W.R., Jones C. \& Tucker W. 2002, \apj, 555, 205

\bibitem[Markevitch, Vikhlinin \& Forman(2002)]{Maxim:aph} Markevitch M., Vikhlinin A. \& 
Forman W.R. 2002, \emph{Matter and energy 
in clusters of galaxies}, ASP Conference Series, Vol. X, Eds. S. Bowyer \& C.-Y. Hwang, 
astro-ph/0208208
\bibitem[Mazzotta et~al(2001)]{Mazzotta:2001} Mazzotta P., Markevitch M., Vikhlinin A.,  
Forman W.R., David L.P. \& VanSpeybroeck L. 2001, \apj, 555, 205
\bibitem[Mazzotta, Edge \& Markevitch(2003)]{Mazzotta:2003} Mazzotta P.,  Edge A.C., 
Markevitch  M. 2003, \apj, 596, 190

\bibitem[Rossetti et~al.(2005; in preparation)]{Rossetti:2005} Rossetti M., Ghizzardi S.,  
Molendi S. \& Finoguenov A. 2005, in preparation

\bibitem[Tittley \& Henriksen(2005)]{TH:2005} Tittley E.R. \& Henriksen M. 2005, \apj, 
618, 227

\bibitem[Vikhlinin, Markevitch \& Murray(2001)]{Vik:2001} Vikhlinin A., Markevitch M. \& 
Murray S.S. 2001, \apj, 551, 160

\end{thebibliography}
\end{document}